# Bessel ultrasonic probe based on conical acoustic lens: simulation study

Xianlin Song [a, #, *], Jianshuang Wei [b, c, #], Lingfang Song [d]

[a] School of Information Engineering, Nanchang University, Nanchang 330031, China;
[b] Britton Chance Center for Biomedical Photonics, Wuhan National Laboratory for Optoelectronics-Huazhong University of Science and Technology, Wuhan 430074, China;
[c] Moe Key Laboratory of Biomedical Photonics of Ministry of Education, Department of Biomedical Engineering, Huazhong University of Science and Technology, Wuhan 430074, China;
[d] Nanchang Normal University, Nanchang 330031, China;
[#] equally contributed to this work;
* songxianlin@ncu.edu.cn

**ABSTRACT**

Ultrasonic transducer is a sensor that realizes the mutual conversion of ultrasonic and electrical signals, and it is widely used in quality inspection, biomedical imaging and other fields. Commonly used ultrasonic transducers have a small detection range and low sensitivity due to the diffraction of sound waves. Focused transducers are used to improve detection sensitivity. Unfortunately, focused transducers have narrow depth of field. Here, we developed a Bessel ultrasonic transducer for large depth of field by using conical acoustic lens. An acoustic lens is attached to a unfocused ultrasonic. And the acoustic lens is a cuboid prism with a concave cone on the bottom, made of fused silica. Similar to an axicon that can generate a Bessel beam, the Bessel ultrasonic transducer can produce nondiffracting Bessel ultrasonic beams. Therefore, extended depth of field with uniformly high resolution and high detection sensitivity can be obtained. We used COMSOL to simulate the transmission of ultrasonic field of the designed conical acoustic lens, and compare it with the spherical focused ultrasonic transducer. The results show that the depth of field of the Bessel ultrasonic transducer is about 8 times that of the conventional spherical focused ultrasonic transducer. And the depth of field of the Bessel ultrasonic transducer can be further adjusted by adjusting the cone angle of the conical acoustic lens. The Bessel ultrasonic transducer will help improve the capabilities of the ultrasound probe and expand its application range. For example, an ultrasonic probe with a large depth of field will expand the imaging depth of photoacoustic microscopy and enhance its ability in non-destructive testing.

## 1. INTRODUCTION

Photoacoustic imaging is a new medical imaging method based on the photoacoustic effect[1]. The photoacoustic effect was proposed in 1880 by Bell. A laser pulse irradiates the sample, the sample absorbs pulse energy, then the temperature rises instantaneously, and adiabatic expansion occurs, which then generates ultrasonic signal. Because the ultrasonic signal is excited by laser pulses, it is called photoacoustic signal. The photoacoustic effect can be successfully applied to biological tissue imaging because of the large differences in the absorption of laser pulses in different biological tissues[2-4]. The photoacoustic signal received by the ultrasonic probe contains the absorption characteristics of biological tissues. The optical absorption distribution can be reconstructed by the reconstruction algorithm.

As a promising tool for biomedical research, optical-resolution photoacoustic microscopy (OR-PAM) is a noninvasive biomedical imaging technique with high-resolution[5]. In the OR-PAM, in order to achieve the coaxial and confocal ultrasonic detection and optical focusing, an acoustic lens is usually used[6]. The acoustic lens can ensure high sensitivity of photoacoustic imaging. The acoustic lens often uses a flat-contact high-frequency probe to glue an irregular prism, which is composed of two right-angle prisms, the two right-angle prisms are placed oppositely, and the gap between the two is filled with silicone oil. The right-angle prism is made of fused silica, and its optical refractive index is close to that of silicone oil. The light beam passes through the prism with a greater transmittance to the bottom of the sample surface.

The acoustic impedance and sound velocity of right-angle prisms are very different from those of silicone oil, so the ultrasonic signal will be reflected when passing through the hypotenuse of the prism, and the contact ultrasonic probe placed sideways detects the ultrasonic signal reflected by the prism. In this way, a coaxial coupling of light and sound can be realized. In order to achieve focused detection, an acoustic lens is glued to the bottom of the prism to achieve highly sensitive detection. At the same time, in order to compensate for the influence of the acoustic lens on the divergence of the beam, a collimating lens is inserted at the top of the prism to ensure that the beam is not affected by the coupling device.

However, the spherical acoustic lens focuses the ultrasound beam strongly, which leads to the limited detection range of the photoacoustic microscopy. In the photoacoustic microscopy, the depth of field (DoF) of the imaging system is mainly determined by the combination of optical focus and ultrasonic focus. There are many ways to improve the optical depth of field, such as using Bessel beam[7], Electrically tunable lens[8]. There are relatively few studies on improving the acoustic depth of field.

In this manuscript, we developed a Bessel ultrasonic transducer for large depth of field by using conical acoustic lens. An acoustic lens is attached to a unfocused ultrasonic. And the acoustic lens is a cuboid prism with a concave cone on the bottom, made of fused silica. Similar to an axicon that can generate a Bessel beam, the Bessel ultrasonic transducer can produce nondiffracting Bessel ultrasonic beams. Therefore, extended depth of field with uniformly high resolution and high detection sensitivity can be obtained. We used COMSOL to simulate the transmission of ultrasonic field of the designed conical acoustic lens, and compare it with the spherical focused ultrasonic transducer. The results show that the depth of field of the Bessel ultrasonic transducer is about 8 times that of the conventional spherical focused ultrasonic transducer. And the depth of field of the Bessel ultrasonic transducer can be further adjusted by adjusting the cone angle of the conical acoustic lens. The Bessel ultrasonic transducer will help improve the capabilities of the ultrasound probe and expand its application range. For example, an ultrasonic probe with a large depth of field will expand the imaging depth of photoacoustic microscopy and enhance its ability in non-destructive testing.

## 2. METHOD

### 2.1 Design of conical acoustic lens

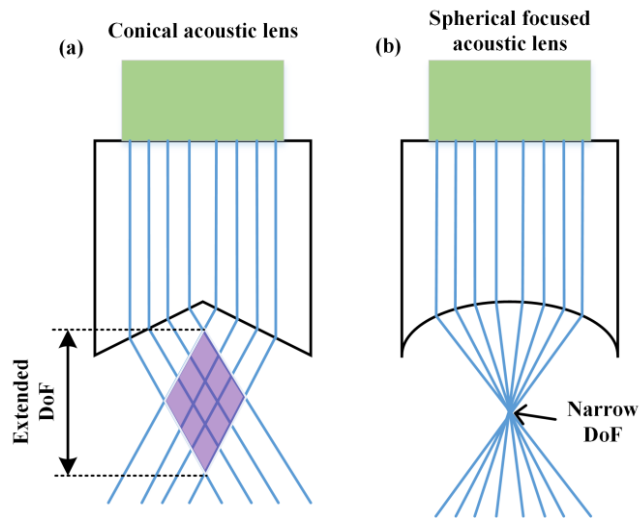

Figure 1. Design of conical acoustic lens. (a) Conical acoustic lens. (b) Spherical focused acoustic lens.

The structure of the conical acoustic lens is shown in Figure 1(a). A flat contact-type high-frequency ultrasonic probe is glued to a prism. The material of the prism is fused silica with a conical concave surface at the bottom. The cone angle can be set according to the needs of the experiment. When a flat-field ultrasonic transducer sends out an ultrasonic signal, since the acoustic impedance of water is much smaller than that of fused silica, the ultrasonic waves transmitted from the conical concave surface of fused silica into the water will deflect inward and generate Bessel sound beams. The Bessel sound beam has the non-diffraction characteristics similar to the Bessel beam, and the size of the sound spot can remain unchanged for a long distance in the axial direction. This area is called the depth of field. Therefore, the introduction of the conical concave surface can make the acoustic lens have a large depth of field. Figure 1(b) is the structure of the spherical focused acoustic lens. since the acoustic impedance of water is much smaller than that of fused silica, the ultrasonic waves transmitted from the conical concave surface of fused silica into the water will deflect inward, therefore, the entire prism is equivalent to a focused acoustic lens, all the sound beams are almost focused on one point, the depth of field is very small.

## 2.2 Simulation using COMSOL

COMSOL Multiphysics is based on the finite element method, which realizes the simulation of real physical phenomena by solving partial differential equations (single field) or partial differential equations (multi-field). It is called "the first truly arbitrary multiphysics" by scientists in the world today. Physics field directly coupled with analysis software". Using mathematical methods to solve real-world physical phenomena, COMSOL Multiphysics realizes highly accurate numerical simulations with efficient calculation performance and outstanding multi-field bidirectional direct coupling analysis capabilities. At present, it has been used in acoustics, biological sciences, chemical reactions, dispersion, electromagnetics, fluid dynamics, fuel cells, earth sciences, heat conduction, microsystems, microwave engineering, optics, photonics, porous media, quantum mechanics, radio frequency, semiconductors, and structures. Mechanics, transmission phenomena, wave propagation and other fields have been widely used.

In this article, the acoustic module of COMSOL is mainly used. The simulation model of the conical acoustic lens is shown in Figure 2(a). The uppermost part is an acoustic lens with a conical concave surface made of fused silica, which can emit ultrasonic signals. The center frequency of the ultrasonic signal is 1 MHz. Below the acoustic lens is water, the role of water is to couple the ultrasonic field. In order to absorb the ultrasonic waves reaching the boundary, a perfect matching layer (PML) is set around the water layer. For comparison, we also built a spherical acoustic lens model, as shown in Figure 2(b).

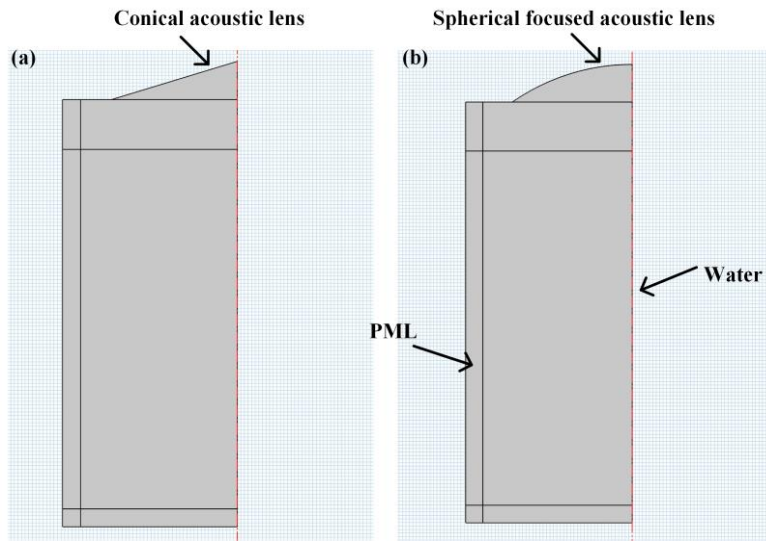

Figure 2. Simulation using COMSOL. (a) Conical acoustic lens. (b) Spherical focused acoustic lens.

# 3. RESULTS

## 3.1 Performance measure

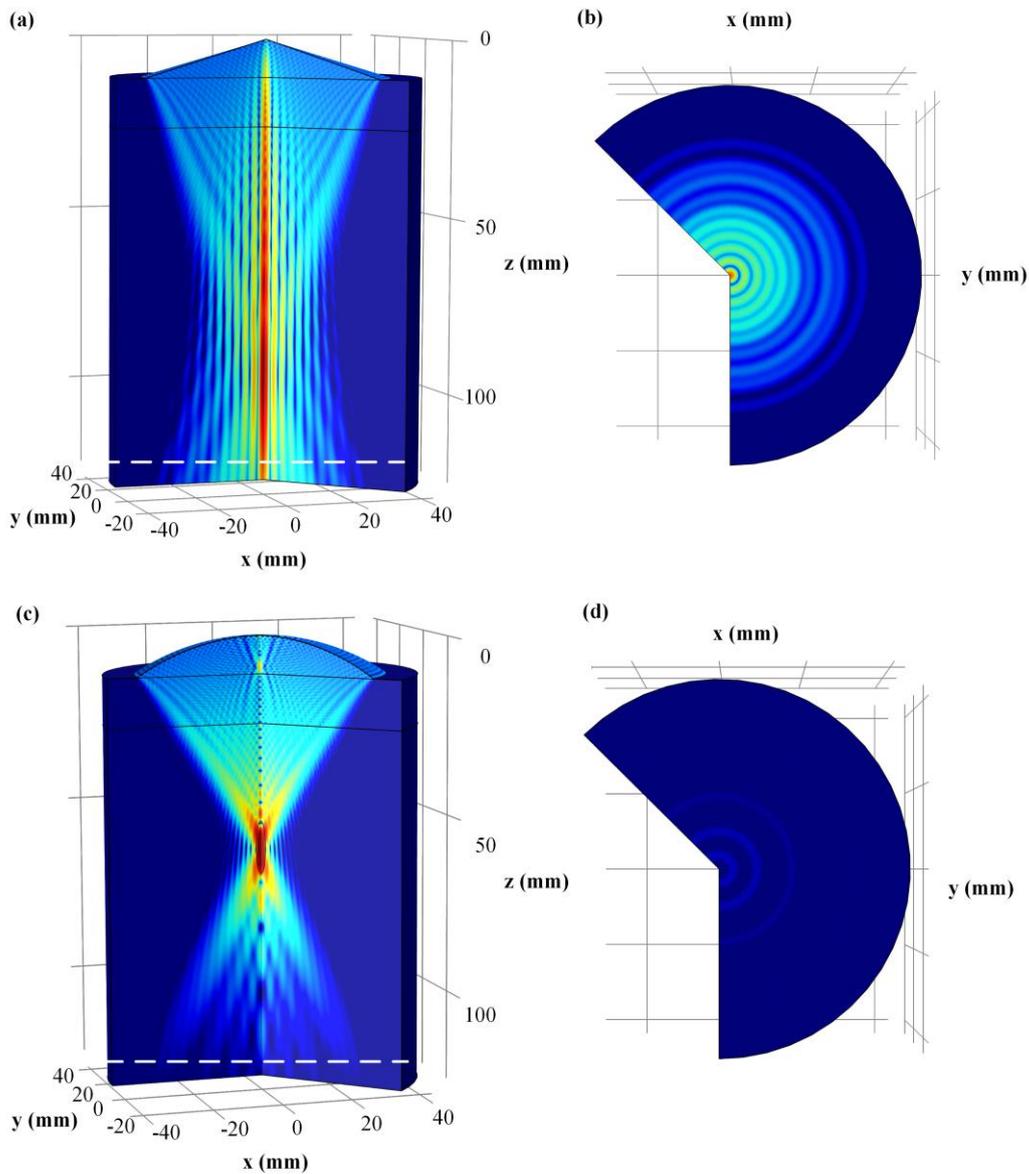

Figure 3. Simulation. (a) The sound field of the conical acoustic lens. (b) The lateral distribution of the sound field indicated by the white dashed line in (a). (c) The sound field of the spherical acoustic lens. (d) The lateral distribution of the sound field indicated by the white dashed line in (c).

Figure 3 shows the results of the model built using COMSOL. The figure shows the field intensity distribution of the sound field produced by the conical acoustic lens and the spherical acoustic lens, respectively. Figure 3(a) shows the sound field of the conical acoustic lens. Because the conical acoustic lens can produce Bessel sound beams, it has non-diffraction characteristics, and the size of the sound spot remains basically unchanged (although some oscillations) over a long axial distance, which means that the conical acoustic lens has large detection depth of field. At the same time, we also see that in the horizontal direction, the field strength of the sound field exhibits a Bessel distribution, as shown in Figure 3(b). In addition to the central main lobe, there are many gradually weakening side lobes beside it. It is because of the existence of these side lobes that the sound beam has self-healing characteristics and keeps its transmission on the axis unchanged. However, for a spherical acoustic lens, all the sound beams are focused in a very narrow range (can be considered as a point), as shown in Figure 3(c), the detection depth of the acoustic lens is very narrow, and the sound spot deteriorates sharply when it is slightly far away from the focus, as shown in Figure 3 (d).

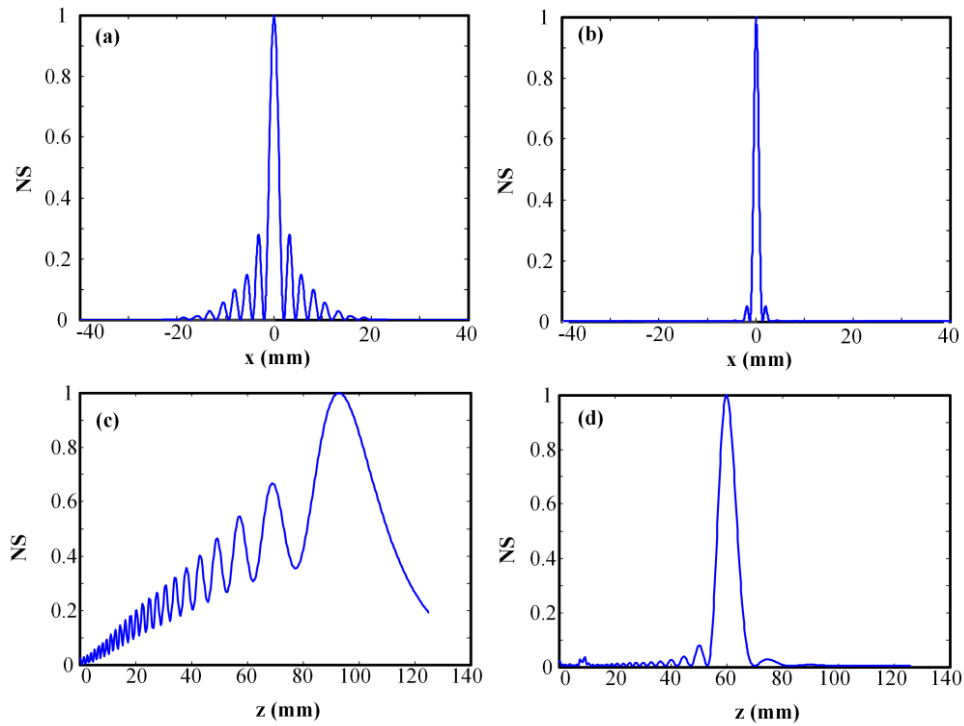

Figure 4. Quantitatively measure the performance of the conical acoustic lens. (a) and (c) are the sound intensity distributions of conical acoustic lens and spherical acoustic lens on the central axis, respectively. (b) and (d) The sound field intensity of the conical acoustic and spherical acoustic lens in the lateral direction, respectively. NS, normalize strength.

In order to quantitatively measure the performance of the conical acoustic lens, we have drawn the axial and lateral sound intensity distributions, as shown in Figure 4. First look at the conical acoustic lens, as shown in Figure 4(a) and Figure 4(b). Figure 4(a) is the sound intensity distribution on the central axis. It can be seen that the sound intensity remains high in the depth range of 0-120 mm. The detection depth of field is defined as the axial distance when the sound field intensity drops to half. Thus, the depth of field of the conical acoustic lens can be calculated to be approximately 52 mm. For the spherical acoustic lens, the measured depth of field is approximately 7 mm, as shown in Figure 4 (c). The depth of field of the conical acoustic lens is ~8 times that of the spherical acoustic lens. In the lateral direction, the sound field intensity of the conical acoustic lens shows a zero-order Bessel distribution, and the full width at half maximum of the center lobe is about 2 mm, as shown in Figure 4 (b). However, for a spherical acoustic lens, the sound spot at the focal point appears as Gaussian distribution, full width at half maximum is 1.3 mm, slightly smaller than the sound spot size of a conical acoustic lens.

## 4. CONCLUSION

We developed a Bessel ultrasonic transducer for large depth of field by using conical acoustic lens. An acoustic lens is attached to a unfocused ultrasonic. And the acoustic lens is a cuboid prism with a concave cone on the bottom. The Bessel ultrasonic transducer can produce nondiffracting Bessel ultrasonic beams. Therefore, extended depth of field with high detection sensitivity can be obtained. The transmission of ultrasonic field of the designed conical acoustic lens was simulated, the results show that the depth of field of the Bessel ultrasonic transducer is about 8 times that of the spherical focused ultrasonic transducer.